\documentclass[twocolumn,prd,aps,nofootinbib,showpacs]{revtex4}
\usepackage{amsmath,amssymb,graphicx,bm}

\begin{document}
\bibliographystyle{apsrev}

\newcommand{\be}{\begin{eqnarray}}
\newcommand{\ee}{\end{eqnarray}}

\title{Instanton Contribution to the Pion Electro-Magnetic Formfactor
at $Q^2 > 1 \, \text{GeV}^2$}
\author{P. Faccioli, A. Schwenk, and E.V. Shuryak}
\affiliation{Department of Physics and Astronomy,
State University of New York, Stony Brook, N.Y. 11794-3800}
\date{\today}

\begin{abstract}
We study the effects of instantons on the charged pion electro-magnetic
formfactor at intermediate momenta. In the Single Instanton
Approximation (SIA), we predict the pion formfactor in the kinematic
region $Q^2 = 2-15 \, \text{GeV}^2$. By developing the
calculation in a mixed time-momentum representation, it is possible to
maximally reduce the model dependence and to calculate the formfactor
directly. We find the intriguing result that the SIA calculation
coincides with the vector dominance monopole form, up to surprisingly
high momentum transfer $Q^2 \sim 10 \, \text{GeV}^2$. This suggests
that vector dominance for the pion holds beyond low energy nuclear physics.
\end{abstract}


\maketitle

\section{Introduction}

Bridging the gap between the non-perturbative and the perturbative 
sector of QCD is a central step toward our understanding of the
strong interaction. In this context, the electro-magnetic
formfactor of the charged pion 
$F_\pi$ is of great interest. It is, at low momenta,
extremely well reproduced by the vector dominance model. In addition,
at very high momentum transfer, it is constrained by 
perturbative (p)QCD predictions (for a review on hadronic 
formfactors, see e.g.~\cite{sterman}). The asymptotic behavior for
large space-like momentum transfer, $Q^2 = - (p-p')^2 > 0$,
is derived in a closed form in perturbation
theory~\cite{pQCDpionchern,pQCDpionrad,pQCDpionbrod},
\be
\label{asyff}
Q^2 \, F_\pi(Q^2) \, \stackrel{Q^2\to\infty}{=} \, 16 \: \pi \: f_\pi^2 \:
\alpha_s(Q) ,
\ee
where $f_\pi = 92.4 \, \text{MeV}$ denotes the pion decay constant.

\begin{figure}[h|t]
\includegraphics[scale=0.32,clip=]{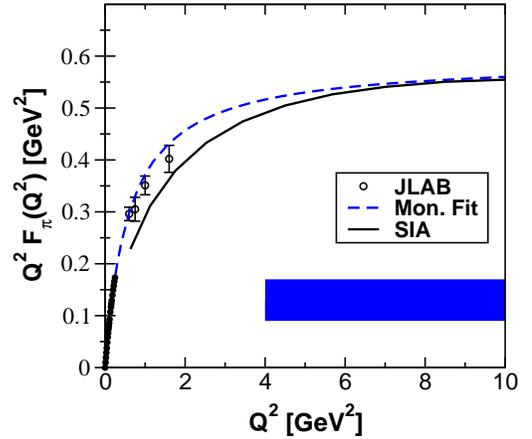}
\caption{The recent JLAB data for $Q^2 \, F_\pi(Q^2)$ in comparison with
the asymptotic pQCD prediction (thick bar, for a typical $\alpha_s
\approx 0.2-0.4$ in Eq.~(\ref{asyff})), the monopole fit
(dashed line), and our SIA calculation (solid line). The SIA
calculation is not reliable below $Q^2 \sim 1 \, \text{GeV}^2$.
The solid circles denote the SLAC data~\cite{SLAC}.}
\label{data1}
\end{figure}

A comparison of the asymptotic behavior and the experimental data
determines the momentum scale where the perturbative regime of QCD is 
reached. Recently, the charged pion formfactor has been
measured very accurately at momentum transfers $0.6 \, \text{GeV}^2 <
Q^2 < 1.6 \, \text{GeV}^2$ by the Jefferson Laboratory (JLAB) $F_\pi$
collaboration~\cite{jlab} and lead to quite surprising results. 
Not only are the data at highest experimentally accessible momenta still
very far from the asymptotic limit, but the trend is away from
the pQCD prediction (see Fig.~\ref{data1}).

Moreover, it is quite remarkable that the data are still completely
consistent with the vector dominance monopole fit,
\be
\label{monopole}
F_{\pi,\text{mon.}} (Q^2)= \frac{M_\rho^2}{M_\rho^2+Q^2} , \quad
M_\rho = 770 \, \text{MeV},
\ee
at relatively high momentum transfer ($Q^2 \approx 1 \, \text{GeV}^2$).
Clearly, the measurements currently undertaken at JLAB in the region
$0.5 \, \text{GeV}^2 \lesssim  Q^2 \lesssim 6 \, \text{GeV}^2$ are
very much needed and will provide information about whether the
perturbative region is reached at that scale.

The charged pion formfactor has attracted a lot of attention from the
theoretical side~\cite{FFtheory1,FFtheory2,FFtheory3,FFtheory4,FFtheory5,FFtheory6,FFtheory7}.
Despite this debate, we feel that
there are still a number of open question.
What are the leading non-perturbative effects responsible for 
the deviation from pQCD, Eq.~(\ref{asyff}), at intermediate momenta? 
Where can we expect the transition to pQCD, is it within
experimentally accessible momentum transfers? Is there a
\emph{microscopic} explanation of the success of the monopole form? Can
it be justified at so large momenta, where the vector dominance model
should be inadequate?

In this work we suggest answers to these questions. In particular, we study
the effects of the leading (i.e. zero-mode) interaction of light
quarks with the intense classical vacuum fields (instantons) on the pion
formfactor, for $Q^2 > 1 \, \text{GeV}^2$. We show that, in this kinematic
region, the pion formfactor is dominated by the interaction of quarks
with a single instanton.

The pion plays a special role among the hadrons. Not only is it nearly massless
(which is explained by the Goldstone theorem), but unusually compact
as well. Phenomenologically, this is seen e.g., from the rather large
electro-magnetic mass splittings between the charged and neutral pion.
Theoretically, it was explained by studying instanton-induced
effects on the two-point correlators, see~\cite{shuryakrev} for a review. 
In the pion channel (as well as other scalar and pseudoscalar
channels) the instanton contribution to the two-point correlator
can be represented by the zero-mode terms in the quark 
propagators, and as a result the effect is enhanced 
(relative to e.g., the vector or axial channels) by a factor
$1/(m^\star \rho)^2$, where $\rho$ denotes the typical instanton size, and 
$m^\star$ is an effective parameter with the dimension of a mass, 
defined and discussed in detail in~\cite{sia}. Numerically, such an enhancement
factor is about 30, and parametrically it is the inverse diluteness of 
the instanton ensemble. Due to the presence of such a large factor, 
instanton-induced forces become dominant in the pion pseudoscalar correlator, 
starting from rather small distances. 
The same factor is present in the instanton contribution to the three-point 
function, which is related to the pion formfactor. 
Although, the final results we obtain are independent of the
value of $m^\star$, the window of applicability of our method 
does depend on it.

This feature, however, is not generically related to the pion itself
and depends on the particular three-point function under
investigation. For example, 
the enhancement is absent, when one considers the pion contribution to
the axial correlator. Similarly, there is no such enhancement of the $\gamma
\gamma^\star \pi^0$ neutral pion transition formfactor. The relevant
instanton effects for this process are not due to (enhanced) zero
modes, but are either related to non-zero mode propagators in 
the instanton background or to multi-instanton effects, which are
suppressed by the instanton diluteness. This conclusion is nicely
supported by  recent CLEO measurements of this formfactor, which 
indeed show that the asymptotic pQCD regime is reached much earlier, 
at $Q^2 \sim 2 \, \text{GeV}^2$~\cite{CLEO}.

The first calculation of the pion formfactor in the SIA
was  performed by Forkel and Nielsen~\cite{forkel}, who
complemented QCD sum rules by the instanton-induced
contribution. In such an approach, however, a model
description of the continuum of excitations with the quantum numbers
of the pion is needed, in order to connect the electro-magnetic three-point
function to the pion formfactor~\footnote{Notice that in~\cite{forkel}
Forkel and Nielsen used the numerical value for the effective parameter 
$m^\star$, which was derived in the mean field
approximation~\cite{shuryak82}. However
later a significantly smaller value (and larger enhancement) has been
derived from numerical simulations of the Instanton
Liquid Model~\cite{sia}.}. In order to avoid such additional model dependence, 
Blotz and Shuryak proposed a different approach~\cite{blotz}, in which
large-sized three-point functions obtained both from simulations in
the Instanton Liquid Model (ILM) and from its spectral decomposition (which
depends on the pion formfactor) were compared \emph{directly in
coordinate space}. This analysis revealed that, at large distances,
the ILM results were completely consistent with the monopole
fit.

In this work we follow a different approach, which is based on the mixed
time-momentum representation widely used in lattice calculations
(see e.g.,~\cite{draper89}). We  show that this way it is possible
to isolate the pion pole and yet make predictions in momentum space
(and therefore compare directly with experiments), without any model
of the continuum contribution.

This approach has several other advantages. Firstly, the model
dependence is maximally reduced: Our results depend only on one
dimensional parameter, the average instanton size, $\bar{\rho} \approx
1/3 \, \text{fm}$, which was proposed many years ago~\cite{shuryak82}
and checked a number of times by phenomenological
studies~\cite{shuryakrev} as well as lattice
simulations~\cite{smithandteper,hasenfratz,deForcrand,chu94,Ivanenko}. 
The calculation is independent of the
properties of the instanton liquid ensemble such
as the instanton-instanton interaction and the instanton density.
Moreover, we will argue that the physical content of the calculation
is more transparent and the dynamical effects of the instantons on the
propagation of the quarks can be seen naturally.

The central prediction following from our SIA calculation is shown in
Fig.~\ref{data1} in comparison to the recent JLAB measurements.

The main physical point of our analysis is that, if the strongly
attractive 't Hooft interaction is taken into account, the 
asymptotic perturbative regime of the pion formfactor is reached much later
than in other models, which do not include this force. Such a
qualitative difference implies that the upcoming JLAB
measurements will provide an important test of instanton
forces in the pion.
We also find the very intriguing result that the instanton 
contribution to the formfactor is completely consistent with the 
monopole form at intermediate momentum transfers, $2 \, \text{GeV}^2 <
Q^2 \lesssim 10 \, \text{GeV}^2$, where the vector dominance model has no
justification. It is a first microscopic study showing that all
other resonances with quantum numbers of the $\rho$ (except the $\rho$
itself) are not seen in the pion formfactor.
For large momentum transfers, $Q^2 > 20 \,
\text{GeV}^2$, our SIA breaks down, as it is necessary to increase the
distances in order to isolate the pion ground state. At these
needed distances, however, the correlation functions will become
sensitive to multi-instanton effects.

\section{Instanton Contribution to the Formfactor}

We consider the spatial Fourier transforms of the Euclidean
three-point function and two-point function, 
\be
\label{3p}
G_\mu (t,{\bf p}+{\bf q};-t,{\bf p}) = \int d^3{\bf x} \,
d^3{\bf y} \, e^{-i \, {\bf p} \cdot {\bf x} + i \,
({\bf p}+{\bf q}) \cdot {\bf y}} \nonumber \\[0.5mm]
\times \langle 0 | \, j_5(t,{\bf y}) \, J_\mu(0,{\bf 0}) \,
j_5^\dagger(-t,{\bf x}) \, | 0 \rangle ,
\ee
\be
\label{2p}
G(2 t,{\bf p}) = \int d^3{\bf x} \,
e^{i \, {\bf p} \cdot {\bf x}} \, \langle 0 | \,
j_5(t,{\bf x}) \, j_5^\dagger(-t,{\bf 0}) \, | 0 \rangle ,
\ee  
where the pseudo-scalar current $j_5(x)=\bar{u}(x) \, \gamma_5 \,
d(x)$ excites states with the quantum numbers of the pion and
$J_\mu(0)$ denotes the electro-magnetic current operator.
In the large $t$ limit (at fixed momenta), 
both correlation functions are dominated by the
pion pole contribution and the ratio of the three-point function to the 
two-point function becomes proportional to the pion
formfactor~\cite{draper89}. In the Breit frame, ${\bf p} = - {\bf
q}/2$ and $Q^2 = {\bf q}^2$, one has simply
\be
\label{FF}
\frac{G_4(t,{\bf q}/2; -t,- {\bf q}/2)}
{G(2 t,{\bf q}/2)} \to F_\pi(Q^2) .
\ee
Notice that the LHS of Eq.~(\ref{FF}) should not depend on $t$, for
$t$ large enough. Below we demonstrate that, for the pion, this 
is achieved already for $t \sim 0.6 \, \text{fm}$.

We will evaluate the LHS of Eq.~(\ref{FF}) in the SIA. In principle,
it is not obvious that such an approach is justified, as the mean
Euclidean distance between two instantons in the vacuum is about $1 \,
\text{fm}$. Therefore, if $t \sim 1 \, \text{fm}$, one would 
expect many instanton effects to play a non negligible role. However, 
two of the authors showed that the pion and nucleon three-point
functions, evaluated in the SIA, agree with the results of numerical
simulations in the instanton liquid model up to distances of the order
of $1 \, \text{fm}$~\cite{sia}. Moreover, they found
that the ratio of three- to two-point function is dominated by single
instanton effects to even larger distances~\cite{3ptILM}. 
This result enables us to reliably evaluate
this ratio in the SIA up to Euclidean
distances of $\sim 1.4 \, \text{fm}$. A possible physical
interpretation of this fact is that in order for the scattered parton
to recombine in the same final state (and therefore have an
elastic formfactor), it is sufficient that the parton scatters off a
single instanton during the process.
  
We expect that the main contribution to 
the time-momentum correlation functions comes from distances of the
order of the inverse conjugate momenta, $|{\bf x}| \lesssim 1/|{\bf p}|$ and
$|{\bf y}| \lesssim 1/|{\bf p}+{\bf q}|$. Therefore, in the Breit
frame and with the choice $t \sim 0.7 \, \text{fm}$, one obtains that
the correlation functions under consideration can be evaluated in the
SIA, for momenta $|{\bf q}| \gtrsim 0.6-1 \, \text{GeV}$. We note that
one of the authors applied the same method to evaluate the pion and
nucleon dispersion curves~\cite{faccioli}. Agreement between the SIA
calculation and experiment was found for $t \gtrsim 0.7 \, \text{fm}$
and $|{\bf p}| \gtrsim 1 \, \text{GeV}$.

Having assessed the applicability of our approximation, 
we proceed to the calculation.
The connected three-point and two-point functions are 
given by~\footnote{A discussion
of the contribution from disconnected diagrams can be found in~\cite{blotz}.}
\begin{multline}
G_4(t,{\bf p} + {\bf q};-t,{\bf p}) = (e_u-e_d) \\[1mm]
\times \int d^3{\bf x} \, d^3{\bf y} e^{-i \, {\bf p} \cdot {\bf x} + i \,
({\bf p} + {\bf q}) \cdot {\bf y}} \langle \text{Tr} \bigl\{ \gamma_5 \,
S(t,{\bf y};0,{\bf 0}) \\
\times \gamma_4 \, S(0,{\bf 0};-t,{\bf x}) \gamma_5 \, S(-t,{\bf
x};t,{\bf y}) \bigr\} \rangle ,
\label{w3}
\end{multline}
\begin{multline}
G(2 t,{\bf p}) = \int d^3{\bf x} \, e^{i \, {\bf p} \cdot {\bf x}}
\langle \text{Tr} \bigl\{ \gamma_5 \, S(t,{\bf x};-t,{\bf 0}) \\
\times \gamma_5 \, S(-t,{\bf 0};t,{\bf x}) \bigr\} \rangle ,
\label{w2}
\end{multline}
where $S(y,x)$ denotes the quark propagator in the background of a
gauge field, the $\text{Tr}$ is over spin and color, and the brackets 
$\langle \: \cdot \: \rangle$ denote the average over all gauge field
configurations.

\begin{figure}[h|t]
\includegraphics[scale=0.6,clip=]{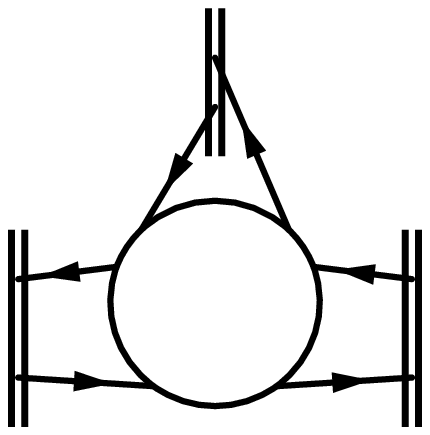}
\hspace*{1cm}
\includegraphics[scale=0.6,clip=]{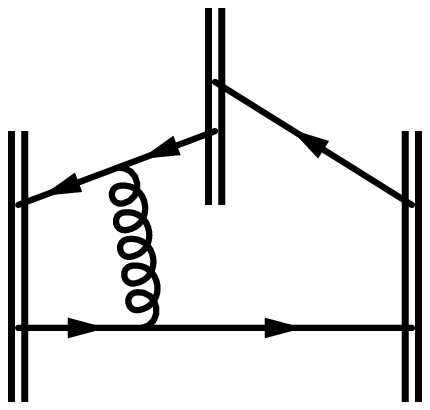} \\
\hspace*{-1.25cm} $-t$ \quad \quad \, $0$ \quad \quad \quad $t$ \quad
\hspace*{2.25cm} (B) \\[1mm]
\hspace*{-4.35cm} (A) \\[2mm]
\includegraphics[scale=0.6,clip=]{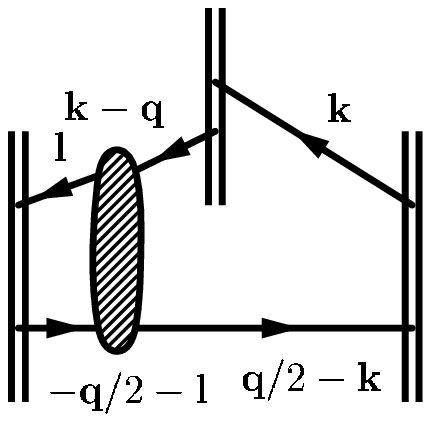}
\hspace*{1.5cm}
\includegraphics[scale=0.6,clip=]{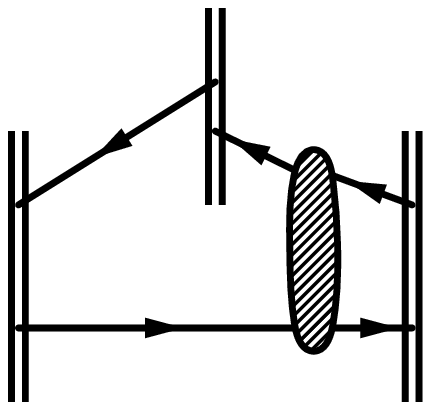} \\[2mm]
(C) \hspace*{3.5cm} (D) \\
\caption{(A): Graphical representation of the W2W three point function,
Eq.~(\ref{ww3p}). The double lined ``walls'' correspond to the spatial Fourier
integration. (B): A particular diagram contributing to (A), in lowest
order pQCD. (C) and (D): The leading non-perturbative contributions to
(A) in the SIA. The dashed ellipse denotes the quark (zero-mode) 't
Hooft interaction. Note that the momentum integrations in
Eqs.~(\ref{ww3p}) and~(\ref{ww2p}) ensure locality at the walls and
the gauge invariance of the calculation.}
\label{wwfig}
\end{figure}

We express Eqs.~(\ref{w3}) and~(\ref{w2}) in terms of ``wall-to-wall''
(W2W) correlators, defined as the spatial Fourier transforms of 
the quark propagators
\be
S(t',{\bf p'};t,{\bf p}) \equiv \int d^3{\bf x} \, d^3{\bf y} \,
e^{i \, {\bf p'} \cdot {\bf y} - i \, {\bf p} \cdot {\bf x}} \, S(y,x)
.
\ee
This is achieved by insertions of appropriate delta functions at each
vertex
\begin{multline}
\label{ww3p}
G_4(t,{\bf p}+{\bf q};-t,{\bf p}) = \int \frac{d^3{\bf k}}{(2\pi)^3}
\int \frac{d^3{\bf l}}{(2\pi)^3} \int \frac{d^3{\bf m}}{(2\pi)^3} \\ 
\times \int \frac{d^3{\bf n}}{(2\pi)^3} \:
\langle \text{Tr} \bigl\{ \gamma_5 \, S(t,{\bf k};0,{\bf m} + {\bf n})
\, \gamma_4 \, S(0,{\bf m};-t,{\bf l}) \\[1mm]
\times \gamma_5 \, S(-t,{\bf l}-{\bf p};t,{\bf k}-{\bf p}-{\bf q}) 
\bigr\} \rangle ,
\end{multline}
\begin{multline}
\label{ww2p}
G(2t,{\bf p}) = \int \frac{d^3{\bf k}}{(2\pi)^3}
\int \frac{d^3{\bf l}}{(2\pi)^3} \int\frac{d^3{\bf m}}{(2\pi)^3} \\[2mm]
\times \langle \text{Tr} \bigl\{ \gamma_5 \, S(t,{\bf k};-t,{\bf l})
\gamma_5 \, S(-t,{\bf m};t,{\bf k}-{\bf p}) \bigl\} \rangle .
\end{multline}
These W2W correlation functions are completely general. Their graphical
representation, as an example for the three-point function,
Eq.~(\ref{ww3p}), is given in Fig.~\ref{wwfig}~(A). 

A model dependence is introduced, when we
evaluate the gauge field average in the SIA. This is achieved by
taking the quark propagators in the background of an instanton and 
averaging over the collective coordinates (instanton position, size
and color orientation). The integration over the
color orientation is trivial, while the integration over the
instanton size is weighted by a distribution function. We use a
distribution for the instanton size which takes into account the
small-size limit. The latter has been calculated by 't~Hooft,
by considering perturbative fluctuations around the instanton solution
\cite{thooftzm1,thooftzm2}. For large-sized instantons, however, the
distribution has to be cut off. We use a Fermi distribution function
to account for the suppression of large-sized instantons, where we fit
the width and the mean instanton size to lattice QCD results (for a
summary on lattice QCD results, see e.g., Negele~\cite{Negele}). 
This leaves for the single instanton density
\be
n(\rho) = \bar{n} \: d_{\text{'t~Hooft}}(\rho) \: \frac{1}{\exp\bigl((\rho -
\bar{\rho})/\sigma \bigr) + 1} ,
\label{latticeQCD}
\ee
where $\bar{n}$ denotes the average instanton density and the QCD
small-size limit is given by 't~Hooft
\begin{multline}
\label{dens}
d_{\text{'t~Hooft}}(\rho) \sim \rho^{N_f-5} \beta_1(\rho)^{2 N_c} \\
\times \exp\bigl\{- \beta_2(\rho) + (2 N_c - \frac{b'}{2 b}) \frac{b'}{2
b \, \beta_1(\rho)} \ln\bigl(\beta_1(\rho)\bigr) \bigr\} .
\end{multline}
Here, $\beta_1(\rho)$ and $\beta_2(\rho)$ denote the one and two-loop beta
functions~\footnote{This result was carried out in the Pauli-Villars
regularization scheme and $\Lambda_{PV}$ is the corresponding scale
parameter. We use $\Lambda_{PV} = 250 \, \text{MeV}.$},
\begin{align}
\beta_1(\rho) &= - b \ln(\rho \, \Lambda_{PV}) , \\
\beta_2(\rho) &= \beta_1(\rho) + \frac{b'}{2 b} \ln\bigl(2 \ln(1/\rho \,
\Lambda_{PV})\bigr) , \\
b &= \frac{11}{3} N_c - \frac{2}{3} N_f , \\
b' &= \frac{34}{3} N_c^2 - \frac{13}{3} N_c N_f + \frac{N_f}{N_c} .
\end{align}
For our analysis, we use a parameterization, which summarizes recent
lattice QCD results~\cite{Negele}, 
$\bar{\rho} = 0.37 \, \text{fm}$ and $\sigma =
0.15 \, \text{fm}$. We will also contrast this size density to a much
simpler model proposed by Shuryak~\cite{shuryak82}, $n(\rho) =
\bar{n} \: \delta(\rho-\bar{\rho})$, where $\bar{\rho} = 1/3 \,
\text{fm}$.

The quark propagator in the field of an instanton was evaluated
exactly in singular gauge~\cite{brown78} and consists of a zero-mode and
a non zero-mode part,
\begin{equation}
\label{full}
S^I(x,y) = S^I_{zm}(x,y) + S^I_{nzm}(x,y) .
\end{equation}
The expression for $S^I_{nzm}$ is quite involved. However, at small
distances, $|x-y| \ll \rho$, and if the instanton is far away,
$|x-z|,|y-z| \gg \rho$, the non zero-mode propagator reduces to the
free one. For this reason, most calculations are carried out in the
zero-mode approximation (ZMA)~\footnote{We note that, at large
distances ($|x-y| \gg \rho$), the zero-mode approximation gives the
same correlation functions as the 't~Hooft Lagrangian.}, in which one approximates the non zero-mode part with a free massless quark propagator,
\be
\label{zeromode}
S^I(x,y) \approx S_0(x,y) + S^I_{zm}(x,y) .
\ee

Let us discuss the accuracy of the ZMA for our calculation. We recall that,
in the momentum range of interest, the contributing point-to-point
correlators in Eqs.~(\ref{3p}) and~(\ref{2p}), are $\sim 1 \,
\text{fm}$ long. It is therefore sufficient to access the accuracy of
the ZMA for our point-to-point correlators at those distances.
In~\cite{sia}, two of the authors have evaluated several three- and
two-point functions in the SIA with ZMA and compared them with the results of
numerical simulations in the instanton liquid model 
(where non-zero mode effects are taken
into account). Very good agreement was observed at distances relevant
for our calculation. This result could either imply that multi-instanton
and non-zero mode effects are individually small at that scale, or
that both effects are not quite small, but that they happen to cancel.

In this approximation, the non-perturbative contributions to 
Eq.~(\ref{ww3p}) are represented in Figs.~\ref{wwfig}~(C) and~(D)
and are straightforward, once the free and zero-mode W2W quark
propagators are evaluated.

The massless free W2W quark propagator is given by
\be
S_0(t',{\bf p'};t,{\bf p})=(2 \pi)^3 \, \delta^{(3)}({\bf p'}-{\bf p})
\, \frac{e^{-|{\bf p}|\,|t'-t|}}{2} \, u_\mu \gamma_\mu , \quad
\label{freeW2W}
\ee
where $u_4=1$ and $u_l=i \, p_l/|{\bf p}|$, for $l=1,2,3$. We note that
the free W2W propagator is proportional to
a delta function due to momentum conservation. The free contribution 
to $G_4$ and $G$ are then given by
\begin{multline} 
\label{G4free}
G_4^{(\text{free})}(t,{\bf p}+{\bf q};-t,{\bf p}) = - \frac{3}{2} \int
\frac{d^3{\bf k}}{(2\pi)^3} \: \biggl( \, \frac{{\bf k} \cdot ({\bf
k}-{\bf q})}{|{\bf k}| \, |{\bf k}-{\bf q}|} \\[1mm]
- \frac{{\bf k} \cdot ({\bf k}-{\bf p}-{\bf q})}{|{\bf k}| \, |{\bf
k}-{\bf p}-{\bf q}|} - \frac{({\bf k}-{\bf q}) \cdot ({\bf k}-{\bf
p}-{\bf q})}{|{\bf k}-{\bf q}| \, |{\bf k}-{\bf p}-{\bf q}|} +1 \biggl) \\[1mm]
\times e^{- t \, (|{\bf k}| + |{\bf k}-{\bf q}| + 2 |{\bf k}-{\bf
p}-{\bf q}|)} ,
\end{multline}
\begin{multline}
\label{Gfree}
G^{(\text{free})}(2t,{\bf p}) = 3 \int \frac{d^3{\bf k}}{(2\pi)^3}
\: \biggl( \, \frac{{\bf k} \cdot ({\bf k}-{\bf p})}{|{\bf k}| \,
|{\bf k}-{\bf p}|} - 1 \biggr) \\[1mm]
\times e^{- 2 t \, (|{\bf k}| + |{\bf k}-{\bf p}|)} .
\end{multline}
The free W2W two-point function can be further simplified,
\begin{equation}
\label{Gfreesimple}
G^{(\text{free})}(2t,{\bf p}) = \frac{3}{(2\pi)^2}
\: \frac{1 + 2t \, |{\bf p}|}{(2t)^3} \: e^{- 2 t \, 
|{\bf p}|} .
\end{equation}

Next, we calculate the zero-mode W2W quark propagator. Since we are
eventually interested in a gauge invariant quantity, and we are working in the
SIA, we can use the instanton solution in regular gauge~\footnote{When 
one considers many instanton effects,
the use of the singular gauge is obligatory, as one desires the
topological charge to be localized. In singular gauge, however, the
expression of the zero-mode W2W quark propagator is more complicated.},
where we find the simple result:
\begin{align}
S^{I(A)}_{zm}(t',{\bf p'};t,{\bf p}) & = \frac{2 \rho^2}{m^\star} \,
f(t',{\bf p'};t,{\bf p}) \, {\bf W}^{I(A)} , \\[1mm]
f(t',{\bf p'};t,{\bf p}) & \equiv e^{i \, ({\bf p'}-{\bf p}) \cdot
{\bf z}} \: K_0 \bigl( |{\bf p'}| \, \sqrt{(t'-z_4)^2+\rho^2} \bigr)
\nonumber \\[1mm]
& \quad \times K_0 \bigl( |{\bf p}| \, \sqrt{(t-z_4)^2+\rho^2}
\bigr) , \\[1mm]
{\bf W}^{I(A)} & \equiv \gamma_\mu \, \gamma_\nu \,
\frac{1\pm\gamma_5}{2} \, \tau_\mu^\mp \,\tau_\nu^\pm ,
\end{align}
where $z_\mu=({\bf z},z_4)$ denotes the instanton position, $m^\star$ is the
effective parameter encoding many-instanton effects defined in~\cite{sia} 
and $\tau_\mu^\pm = ({\bm\tau},\mp i)$ are color matrices. We remark that the 
spatial instanton position appears only coupled to the quark momentum
transfer in the background of the instanton field.
This is intuitive: if the instanton is far away from the
quark, there should be little momentum transfer. Due to this
fact, the integration over the spatial instanton position leads to an
overall momentum conserving delta function. As expected, the momentum
conservation in the SIA is a consequence of the translational invariance of the
spatial instanton position. Moreover, the modified Bessel functions
exponentially suppress momentum transfers much larger than the inverse
size of the instanton, $1/\rho$. This implies that large
momentum transfers are driven by small instantons and directly probe the
small-size limit of the 't~Hooft measure for the tunneling amplitude.
This argument holds for all W2W correlation functions evaluated in the
SIA and was applied by one of the authors to the pion and nucleon
correlators~\cite{faccioli}. It was shown that at sufficiently large
momenta, all the model-dependence in such a semi-classical approach
is removed, and the physical observables can be expressed in terms of
$\Lambda_{\text{QCD}}$. We shall similarly study at what
momentum transfer the pion formfactor becomes sensitive to the
small-sized instantons only.

The explicit evaluations of the zero-mode contributions,
Figs.~\ref{wwfig}~(C) and~(D), lead to
\begin{multline}
\label{resultVL}
G_{4 \, zm}^{(C)} = \int_0^\infty d\rho \int_{-\infty}^\infty d z_4
\int \frac{d^3{\bf k}}{(2 \pi)^3} \int \frac{d^3{\bf l}}{(2 \pi)^3} \:
n(\rho) \\[1mm] 
\times \frac{512 \rho^4}{m^{\star\,2}} \: e^{-|{\bf k}|\,t} \: 
K_0\bigl(|{\bf k}-{\bf q}| \: \xi(z_4)\bigr) \: K_0\bigl(|{\bf q}/2-{\bf
k}| \: \xi(z_4-t)\bigr) \\[1mm]
\times K_0\bigl(|-{\bf l}-{\bf q}/2| \:
\xi(z_4+t)\bigr) \: K_0\bigl(|{\bf l}| \: \xi(z_4+t)\bigr) ,
\end{multline}
\begin{multline}
\label{resultVR}
G_{4 \, zm}^{(D)} = \int_0^\infty d\rho \int_{-\infty}^\infty d z_4 
\int \frac{d^3{\bf k}}{(2 \pi)^3} \int \frac{d^3{\bf l}}{(2 \pi)^3} \:
n(\rho) \\[1mm]
\times \frac{512 \rho^4}{m^{\star\,2}} \: K_0\bigl(|{\bf k}| \:
\xi(z_4-t)\bigr) \: K_0\bigl(|{\bf q}/2-{\bf k}| \: \xi(z_4-t)\bigr)
\: e^{-|{\bf l}|\,t} \\[1mm]
\times  K_0\bigl(|-{\bf l}-{\bf q}/2| \: \xi(z_4+t)\bigr) \:
K_0\bigl(|{\bf l}+{\bf q}| \: \xi(z_4)\bigr) ,
\end{multline}
where $\xi(x) \equiv \sqrt{x^2+\rho^2}$. We have labeled the momenta,
flowing from left to right, for the case of Eq.~(\ref{resultVL}) in
Fig.~\ref{wwfig}~(C). We note that the presence of the
(non-perturbative) instanton-induced interaction results in an
integration over the momentum exchanged by the quarks through the
instanton, similar to the presence of perturbative gluon exchanges
leading to loop integrals, Fig.~\ref{wwfig}~(B). This analogy still
holds when multi-instanton effects are taken into account: each
instanton gives an additional ``loop'' integral over the instanton
induced momentum transfer. Furthermore, we read off the ``Feynman
rules'': each free quark propagator contributes an exponential and
each zero-mode a modified Bessel function. Combined, the W2W
three-point function, Eq~(\ref{ww3p}), in the SIA reads
\be
G_4 = G_4^{(\text{free})} + \: G_{4 \, zm}^{(C)} \: + \: G_{4 \, zm}^{(D)} .
\ee
Time reversal invariance imposes that the contributions from
Figs.~\ref{wwfig}~(C) and~(D) are equal, as can be seen from
Eqs.~(\ref{resultVL}) and~(\ref{resultVR}) by substituting $t \to -t$
and ${\bf q} \to -{\bf q}$ (recall the absolute value in the
exponential of the free W2W propagator, Eq.~(\ref{freeW2W})).
Finally, the W2W two-point function, Eq.~(\ref{ww2p}), in the SIA, is
given by
\begin{multline}
\label{twozm}
\hspace*{-3mm}
G_{zm} = \int_0^\infty d\rho \int_{-\infty}^\infty d z_4
\: n(\rho) \: \frac{\rho^4}{4 \, m^{\star\,2}} \: e^{- |{\bf q}|/2 \:
\bigl(\xi(z_4-2t)+\xi(z_4)\bigr)} \\
\times \frac{(1 +|{\bf q}| \, \xi(z_4-2t)/2)}{\xi(z_4-2t)^3}
\: \frac{(1 +|{\bf q}| \, \xi(z_4)/2)}{\xi(z_4)^3} ,
\end{multline}
and analogously
\be
G = G^{(\text{free})} + \: G_{zm} .
\ee 

The pion formfactor is now readily obtained by means of
Eq.~(\ref{FF}). At the distances we are considering ($t = 0.7 \, 
\text{fm}$), the free contribution to the correlation functions,
$G_4^{(\text{free})}$ and $G^{(\text{free})}$, are at most $30 \%$
corrections (at $|{\bf q}| = 5 \, \text{GeV}$) to the zero-mode three-
and two-point functions. This can be seen by evaluating e.g., the
zero-mode two-point function for large $|{\bf q}|$:
\be
G_{zm} = \int_0^\infty d\rho \: \frac{n(\rho) \, \rho^4}{16 \, 
m^{\star\,2} \, \rho} \: \sqrt{\frac{2\pi}{\xi^5(t)}} \: |{\bf
q}|^{3/2} \: e^{ - |{\bf q}| \: \xi(t) } ,
\ee
by a saddle point analysis of the $z_4$ integration. In comparison to
the free contribution, Eq.~(\ref{Gfreesimple}), this leads to a $30
\%$ correction at $|{\bf q}| = 5 \, \text{GeV}$, for
standard values of the average instanton density $\bar{n} = 1 \,
\text{fm}^{-4}$ and the quark effective mass $m^{\star} = 70 \, \text{MeV}$.
We consider such a correction to be the edge of the window we work
in~\footnote{Note, however, that similar corrections appear in the
numerator and the denominator, and thus the real accuracy of the
formfactor can actually be better.}.
Therefore, we may approximate the formfactor
by~\footnote{We note that the free correlators do not have a pion
pole. They are the leading order $\mathcal{O}(\alpha_s^0)$ continuum
contributions. The relevance of these terms at higher
momentum transfer demands larger times $t$. As discussed, however, then the
SIA breaks down.}
\begin{align}
&\frac{G_4^{(\text{free})}(t,\frac{{\bf q}}{2}; -t,- \frac{{\bf q}}{2}) 
+ G_{4 \, zm}^{(C) + (D)}(t,\frac{{\bf q}}{2}; -t,- \frac{{\bf q}}{2})}
{G^{(\text{free})}(2 t,{\bf q}/2)+ G_{zm}(2 t,{\bf q}/2)} \nonumber \\[1mm]
\simeq & \: \frac{G_{4 \, zm}^{(C) + (D)}(t,\frac{{\bf q}}{2}; -t,-
\frac{{\bf q}}{2})}{G_{zm}(2 t,{\bf q}/2)} \to F_\pi(Q^2) .
\label{3to2}
\end{align}
Consequently, the effective parameter $m^\star$ and the average
instanton density $\bar{n}$, which are multi-instanton induced model
parameters, are irrelevant for the pion formfactor at intermediate
momentum transfer. We stress that, in the approach presented here, 
we do not need to evaluate the pion wave function, since we obtain 
directly the relevant Green functions from the SIA. In other words, the 
pion wave function is implicitly included in our calculation through 
the large time extraction of the pion pole.

\section{Numerical Results}

\begin{figure}[b]
\includegraphics[scale=0.32,clip=]{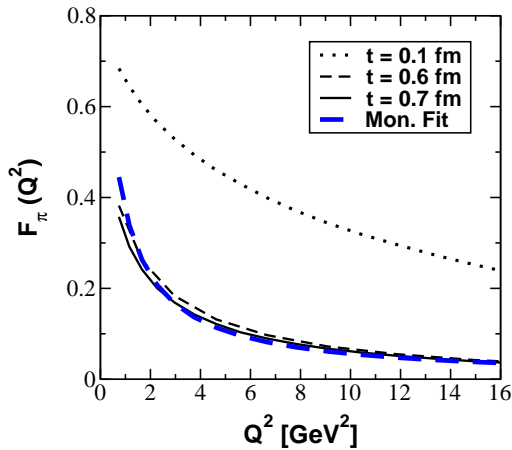}
\caption{The ratio of three- to two-point function, Eq.~(\ref{3to2}),
evaluated in the SIA for different values of $t$ and for simplicity a 
delta function distribution at $\bar{\rho} = 1/3 \, \text{fm}$. At $t
> 0.5 \, \text{fm}$ our results are $t$ independent.} 
\label{tdep}
\end{figure}

Before comparing our results to the 
experimental data, we check that the ratio of
the zero-mode three-point to two-point function, Eq.~(\ref{3to2}), does not
depend on the distance $t$, which ensures that the pion pole
contribution has been successfully isolated. For this purpose, we show
this ratio for several distances $t$ in Fig.~\ref{tdep}. We find that
the pion pole contribution dominates for $t > 0.5 \, \text{fm}$,
in agreement with~\cite{shuryakrev,faccioli}.

Our central result for the pion formfactor is plotted in Fig.~\ref{data1}
in comparison to the JLAB data and the monopole formfactor fitted to
the low $Q^2$ data. We observe that the SIA prediction agrees with the
recent JLAB measurements. In addition, 
it is completely consistent with the monopole form in the
kinematic region accessible to JLAB. These results complement the
analysis of Blotz and Shuryak~\cite{blotz}, where the same agreement 
was found at small momentum transfer. Upcoming measurements at JLAB
will be able to test the single instanton prediction as the microscopic
mechanism at intermediate momentum transfer.

We find that the instanton contribution remains well above the
pQCD scale set by Eq.~(\ref{asyff}) throughout the experimental
region under investigation at JLAB, $Q^2 \lesssim 6 \, \text{GeV}^2$.

As the momentum is increased, one requires larger times in order to
isolate the pion pole from the higher excitations. In sum-rule approaches,
contributions from the continuum are obtained from the free and
perturbative gluon exchange correlation functions, e.g., Fig.~\ref{wwfig}~(B).
Perturbative contributions will never develop a pion pole. Indeed, we
observed that the free correlators, $G^{(\text{free})}_4$ and
$G^{(\text{free})}$, become non negligible for $|{\bf q}| \gtrsim 4-5
\, \text{GeV}$. The presence of such continuum contributions
destroys the time-independence of the ratio of three- to two-point
functions around $t \sim 0.6 \, \text{fm}$, shown in Fig.~\ref{tdep},
and hence the method can no longer be used.

In principle, one could extend the range of validity of our approach by 
increasing the time. However we recall that, when $t$ becomes larger,
multi-instanton effects are significant and the SIA breaks down.
We conclude that our approach is not applicable to the study of the
high momentum transfer region, $Q^2 > 20 \, \text{GeV}^2$.
        
\begin{figure}[b]
\includegraphics[scale=0.6,clip=]{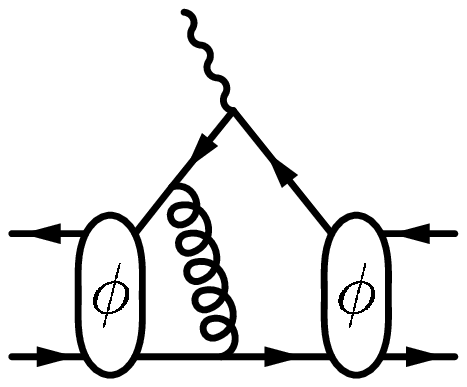}
\hspace*{1cm}
\includegraphics[scale=0.6,clip=]{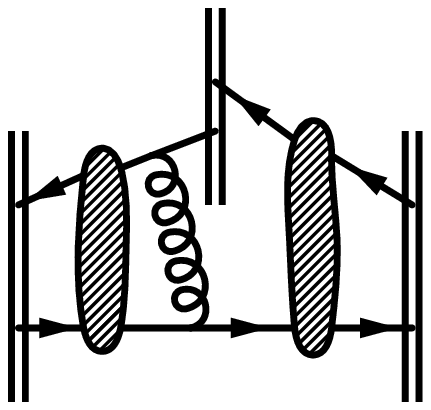} \\[2mm]
(E) \hspace*{3.5cm} (F) \\
\caption{(E): Diagram contributing to the formfactor at asymptotic
momentum transfer in pQCD. The blobs denote the pion wave functions 
$\phi$. (F) A leading order multi-instanton contribution to the W2W
three-point correlator responsible for the transition toward the pQCD 
result.}
\label{wwpQCDvMI}
\end{figure}

Let us now discuss, how one can get the transition to the pQCD limit
from instantons. For large momentum transfer, where counting rules and
factorization are justified, the evaluation of the formfactor 
requires non-perturbative information, encoded in the pion wave
function $\phi$, in addition to pQCD diagrams. This is shown
diagrammatically in Fig.~\ref{wwpQCDvMI}~(E).
In the instanton liquid model, the pion is bound because the
quark and antiquark feel a strong attraction from instantons.
As we argued, at large momentum transfers, one has to incorporate 
multi-instanton effects in order to isolate the pion pole and the
pion wave function. For example, we give a leading order
multi-instanton diagram relevant in the asymptotic limit in 
Fig.~\ref{wwpQCDvMI}~(F).

\begin{figure}[b] 
\includegraphics[scale=0.32,clip=]{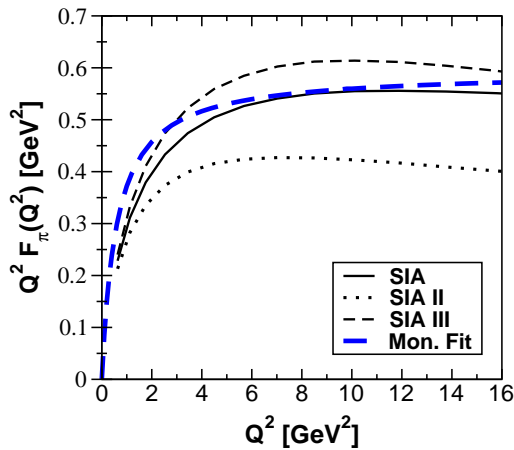}
\caption{The dependence of the pion formfactor $Q^2 F_\pi(Q^2)$ on the
instanton size distribution. The SIA (solid) curve represents a
small-size 't~Hooft distribution with a Fermi distribution cutoff and
lattice QCD parameters. The SIA II (dotted) curve has a different
mean instanton size $\bar{\rho} = 0.47 \, \text{fm}$ (same width) and
the SIA III (dashed) curve is obtained with the simplest delta distribution
$n(\rho) = \bar{n} \, \delta(\rho - 1/3 \, \text{fm})$.}
\label{FFndep}
\end{figure}

Since the pion formfactor may experimentally be rather accurately measured,
it is instructive to ask whether such data may shed some light 
on the instanton size
distribution. In Fig.~\ref{FFndep}, we have plotted the results of 
our theoretical predictions for $Q^2 \, F_\pi(Q^2)$ obtained for
different cases of $n(\rho)$. We contrast the simplest size 
distribution~\cite{shuryak82}, $n(\rho) \sim \delta(\rho-1/3 \,
\text{fm})$, to the results obtained from a lattice QCD
parameterization. We notice that the presence of tails in the size
distribution introduces only small corrections to the
formfactor. Therefore, we conclude that the simplest distribution
$n(\rho) \sim \delta(\rho-1/3 \, \text{fm})$ indeed captures the
relevant features for the pion formfactor at intermediate momentum transfer. 
We observe that our result becomes closer to the perturbative limit, 
if the average instanton size is larger or possibly if there is 
an asymmetric tail toward larger-sized instantons in the distribution.

Next, we compare the results from distributions which have the
same small-size limit, but cut off at different instanton
sizes, $\bar{\rho} = 0.37 \, \text{fm}$ and $0.47 \, \text{fm}$.
We observe that, throughout the entire kinematic region we 
have considered, our predictions are quite sensitive to the 
average instanton size. This implies that the asymptotic region, where
the pion formfactor probes the small-size 't~Hooft behavior, is
not reached within the window, where our approach is justified. 

Summarizing, we have shown that the presence of enhanced instanton-induced
forces constitutes a large and clearly dominant contribution to the charged
pion formfactor. The latter is so large that the perturbative regime
will not be reached experimentally. This nicely contrasts the situation 
for the $\gamma \gamma^\star \pi^0$ neutral pion transition formfactor, 
where the asymptotic pQCD regime is reached much earlier, at $Q^2 \sim
2 \, \text{GeV}^2$~\cite{CLEO}. This striking difference is explained by
instanton arguments as well, because the two corresponding three-point
functions have a different chiral structure. Physically, this is the same
reason why the vector and axial channels have a rather strong ``Zweig''
rule, forbidding flavor mixing, while for the pseudoscalars such a
mixing is very strong.

\section{Conclusions}

We have presented the first calculation of the instanton contribution
to the charged pion electro-magnetic formfactor in the momentum range
$Q^2 > 1 \, \text{GeV}^2$. Technically, it is based on the ratio of
three- to two-point correlators, in which the enhancement comes from
the zero-mode quark W2W propagators. Because it is based on a 
representation in momentum space, the framework developed presents 
several quite direct analogies with ordinary perturbative
calculus. 

Our calculation shows that, when the non-perturbative instanton-induced 
forces are taken into account, the charged pion formfactor 
remains much larger than the asymptotic perturbative prediction, 
throughout the entire experimentally accessible region of JLAB. 
This result is in contrast to other model calculations, which 
do not include such a 
force~\cite{FFtheory1,FFtheory2,FFtheory3,FFtheory4,FFtheory5,FFtheory6,FFtheory7}.
Clearly, the upcoming JLAB measurements will provide a unique opportunity
to test the role played by the instanton-induced 't~Hooft 
interaction in hadronic physics. 

Moreover, we found  that the SIA prediction coincides with the 
monopole form up to $Q^2 \sim 10 \, \text{GeV}^2$. 
We emphasize that this does not represent a test of the validity of our model, 
since the monopole fit is a phenomenological parametrization
exclusively of the low energy data.
Nevertheless, such an agreement is interesting, because
it suggests that the widely used vector dominance model for the pion is 
in fact a dynamic property of the QCD vacuum, and that its validity 
extends into the intermediate momentum regime.

We finally studied whether our results depend on the details of the
instanton size distribution. We found that the pion
formfactor is determined by the average instanton size up to rather
large $Q^2$ and is insensitive to the details of the size distributions.

In addition, an application of the same method to the analysis of the proton
electric formfactor has been carried out and leads to a dipole
formfactor at intermediate momentum transfers~\cite{protonFF}.

\begin{acknowledgments}
We would like to thank Thomas Sch\"afer and George Sterman 
for interesting discussions. The work is supported by the US DOE grant
No. DE-FG02-88ER40388.
\end{acknowledgments}

\end{document}